\documentclass[aps,prb,twocolumn,floatfix,showpacs,superscriptaddress]{revtex4-2}
\usepackage{graphics}
\usepackage{epsfig}
\usepackage{times}
\usepackage{bm}
\usepackage{braket}
\usepackage{color}

\usepackage{amsmath}	
\begin{document}
\title{Nontrivial three-sublattice magnetization in the easy-axis spin-$1/2$ XXZ antiferromagnet on the triangular lattice}
\author{Masahiro Kadosawa}
\affiliation{Department of Physics, Chiba University, Chiba 263-8522, Japan}
\author{Masaaki Nakamura}
\affiliation{Department of Physics, Ehime University, Ehime 790-8577, Japan}
\author{Yukinori Ohta}
\affiliation{Department of Physics, Chiba University, Chiba 263-8522, Japan}
\author{Satoshi Nishimoto}\email{s.nishimoto@ifw-dresden.de}
\affiliation{Department of Physics, Technical University Dresden, 01069 Dresden, Germany}
\affiliation{Institute for Theoretical Solid State Physics, IFW Dresden, 01069 Dresden, Germany}

\date{\today}

\begin{abstract}
	We investigate the ground-state magnetic structure of the spin-$1/2$ XXZ antiferromagnet on the triangular lattice in the easy-axis regime using the density-matrix renormalization group. By applying spiral boundary conditions, we exactly map finite $L\times L$ clusters onto one-dimensional chains while avoiding the spatial anisotropy inherent in cylindrical geometries. From symmetry-broken local magnetization profiles, we extract the three-sublattice moments and track their evolution with anisotropy. At the isotropic point, we obtain a positive sublattice moment of $0.217(3)$, consistent with previous numerical estimates. In the easy-axis regime ($\Delta=J_z/J_\perp>1$), the ordered moments remain close to a Y-like zero-magnetization three-sublattice state, whose $z$-component pattern is of the form $(2m,-m,-m)$, over a broad range of $\Delta$.
	Extrapolation in $1/\Delta$ shows that the positive sublattice moment stays well below the classical saturation value $1/2$, approaching $0.419(7)$ as $\Delta\to\infty$, while the magnitude of the negative sublattice moment approaches $0.209(4)$. We further compare the energies of the Y state and the up-down-down state and find that the Y state is favored at zero field. Independent thermodynamic-limit energy calculations, performed without assuming any particular ordered pattern, yield an energy consistent with the Y-state solution. These results show that the easy-axis ground state does not simply cross over to a trivially saturated collinear Ising state, but instead remains a nontrivial three-sublattice ordered state selected from the macroscopically degenerate Ising manifold by quantum fluctuations.
\end{abstract}

\maketitle

\section{Introduction}

Frustrated quantum magnets on the triangular lattice provide a paradigmatic setting for exploring the interplay among geometric frustration, exchange anisotropy, and quantum fluctuations~\cite{Ramirez1994, Balents2010,Lacroix2011}. A central example is the spin-$1/2$ XXZ Heisenberg antiferromagnet on the triangular lattice, which interpolates between the isotropic Heisenberg limit, where the ground state exhibits three-sublattice $120^\circ$ magnetic order, and the easy-axis regime, where the system acquires an increasingly Ising-like character~\cite{Miyashita1985,Sellmann2015}. Here, the easy-axis regime refers to $\Delta=J_z/J_\perp>1$, where $J_z$ and $J_\perp$ denote the longitudinal and transverse exchange couplings, respectively. In the classical spin picture, the Y state discussed below is a zero-magnetization three-sublattice coplanar state in which one sublattice spin points predominantly along the easy axis while the other two are tilted symmetrically. Beyond its significance for frustrated magnetism, this model is also closely related to hard-core bosons on the triangular lattice, providing a common framework for discussing magnetic order, density-wave order, and supersolid behavior~\cite{Wessel2005,Heidarian2005,Boninsegni2005,Melko2005}.

In the isotropic limit, the existence of three-sublattice order is well established, although quantitative estimates of the ordered moment still show noticeable method dependence~\cite{Huse1988,Bernu1992,Bernu1994,Capriotti1999,Zheng2006,White2007,Li2022,Huang2024}. Experimentally, the spin-$1/2$ triangular-lattice antiferromagnet Ba$_3$CoSb$_2$O$_9$ has provided a canonical realization of nearly ideal triangular-lattice physics~\cite{Shirata2012}. The easy-axis regime has recently attracted renewed attention in connection with triangular-lattice antiferromagnets such as Na$_2$BaCo(PO$_4$)$_2$~\cite{Gao2022,Sheng2022} and K$_2$Co(SeO$_3$)$_2$~\cite{Zhu2025}, where spin-supersolid behavior and related three-sublattice ordering phenomena have been discussed. This experimental context makes it particularly timely to clarify how the magnetic structure of the minimal spin-$1/2$ XXZ model evolves from the isotropic point toward the large-$\Delta$ regime. Related DMRG work on the extended triangular-lattice $J_1$--$J_2$ XXZ model has also addressed the Y/supersolid order parameters and the absence of a net $S^z$ magnetization in the easy-axis regime~\cite{Cesar2025}.

For bipartite lattices such as the square and honeycomb lattices, the easy-axis limit is conceptually straightforward: as the XXZ anisotropy increases, the ground state evolves toward a N\'eel-type ordered state, and the sublattice magnetization approaches its classical saturation value~\cite{Anderson1952,Sachdev2023,kadosawa2023-2,Kadosawa2024}. On the triangular lattice, however, the situation is fundamentally different. In the strict Ising limit, the nearest-neighbor antiferromagnetic Ising model is macroscopically degenerate at zero field, and no unique long-range-ordered ground state is selected~\cite{Wannier1950}. As a result, the large-anisotropy regime of the triangular-lattice XXZ model cannot be understood \emph{a priori} as a trivial collinear Ising state.

The key question is therefore how the easy-axis ground state approaches the Ising limit. More specifically, it is important to determine whether increasing anisotropy drives the system toward a nearly collinear up-down-down-type arrangement, whether a distinct three-sublattice structure persists even for very large $\Delta$, or whether the macroscopic degeneracy of the Ising manifold leaves the nature of the large-$\Delta$ ground state nontrivial \emph{a priori}. This issue is tied to the difference between the strict Ising limit and the large-but-finite-anisotropy regime. While the former is governed by a massively degenerate classical manifold~\cite{Wannier1950}, the latter allows the transverse exchange to lift this degeneracy through a mechanism akin to quantum order-by-disorder~\cite{Villain1980,Shender1982} and thereby select a specific quantum ground state~\cite{Sellmann2015,Sen2008}. To the best of our knowledge, however, an explicit estimate of the asymptotic three-sublattice magnetization in the $\Delta\to\infty$ limit has not yet been reported.

In this work, we investigate the ground-state properties of the spin-$1/2$ XXZ antiferromagnet on the triangular lattice at zero magnetic field, focusing on the evolution of the three-sublattice magnetization in the easy-axis regime. Using the density-matrix renormalization group (DMRG) with spiral boundary conditions (SBC)~\cite{Nakamura2021,kadosawa2023-1}, which avoid the spatial anisotropy inherent in cylindrical geometries, we determine the sublattice moments from the local magnetization profile and obtain an explicit estimate of their asymptotic behavior toward the Ising limit. We also compare these results with thermodynamic-limit calculations performed without pinning fields and without assuming any particular ordered pattern. Our results show that the $z$ component of the ordered moment is well described by a three-sublattice pattern of the form $(2m,-m,-m)$, while the largest sublattice moment remains well below the classical saturation value even in the large-$\Delta$ limit. In addition, the thermodynamic-limit energy obtained without pinning fields agrees with the Y-state solution. Taken together, these results indicate that the easy-axis ground state does not simply cross over to a trivially saturated collinear Ising state.

This perspective is also relevant to recent experiments on easy-axis triangular antiferromagnets. In particular, the zero-field ordered state of K$_2$Co(SeO$_3$)$_2$ has been identified as a three-sublattice Y-type state with zero net magnetization, emerging from the Wannier-degenerate Ising manifold~\cite{Zhu2025,Wannier1950}. The present study provides a microscopic basis for understanding such behavior within the minimal spin-$1/2$ XXZ model.

The remainder of this paper is organized as follows. In Sec.~\ref{sec:model}, we introduce the model. In Sec.~\ref{sec:method}, we describe the numerical method, including the SBC construction and the DMRG procedure. In Sec.~\ref{sec:results}, we present the anisotropy dependence of the three-sublattice magnetization and compare the energies of the Y and up-down-down states. Finally, Sec.~\ref{sec:conclusion} summarizes the results.

\section{Model}
\label{sec:model}

We consider the spin-$1/2$ XXZ Heisenberg antiferromagnet on the triangular lattice,
\begin{equation}
	H=\sum_{\langle ij\rangle}
	\left(
	S_i^x S_j^x + S_i^y S_j^y + \Delta S_i^z S_j^z
	\right),
	\label{eq:Hamiltonian}
\end{equation}
where $S_i^\gamma$ ($\gamma=x,y,z$) denotes a spin-$1/2$ operator at site $i$, $\Delta$ is the exchange anisotropy, and $\langle ij\rangle$ runs over nearest-neighbor bonds.

The ground-state properties of Eq.~(\ref{eq:Hamiltonian}) depend on the exchange anisotropy $\Delta$. In particular, on the triangular lattice geometric frustration makes the approach to the easy-axis Ising limit nontrivial. In the easy-plane region $-1 < \Delta < 1$, the ground state is generally characterized by a three-sublattice noncollinear magnetic structure continuously connected to the $120^\circ$ state at the isotropic point $\Delta = 1$~\cite{Miyashita1985,Sellmann2015}. Although the present work focuses on the easy-axis regime, special exactly solvable points in the easy-plane regime have also been discussed in previous work~\cite{Pal2021}. In the easy-axis region $\Delta > 1$, the model approaches the nearest-neighbor antiferromagnetic Ising model on the triangular lattice, whose zero-field ground-state manifold is macroscopically degenerate in the strict Ising limit~\cite{Wannier1950}. For finite but large $\Delta$, however, the transverse exchange terms lift this degeneracy and can stabilize an ordered state. For $\Delta < -1$, the ground state is ferromagnetic with spins fully polarized along the $z$ direction. Thus, unlike the square- and honeycomb-lattice cases, the large-$\Delta$ limit of the triangular-lattice XXZ model cannot be understood simply in terms of a unique classical N\'eel state.

\begin{figure}[tbh]
	\centering
	\includegraphics[width=0.7\linewidth]{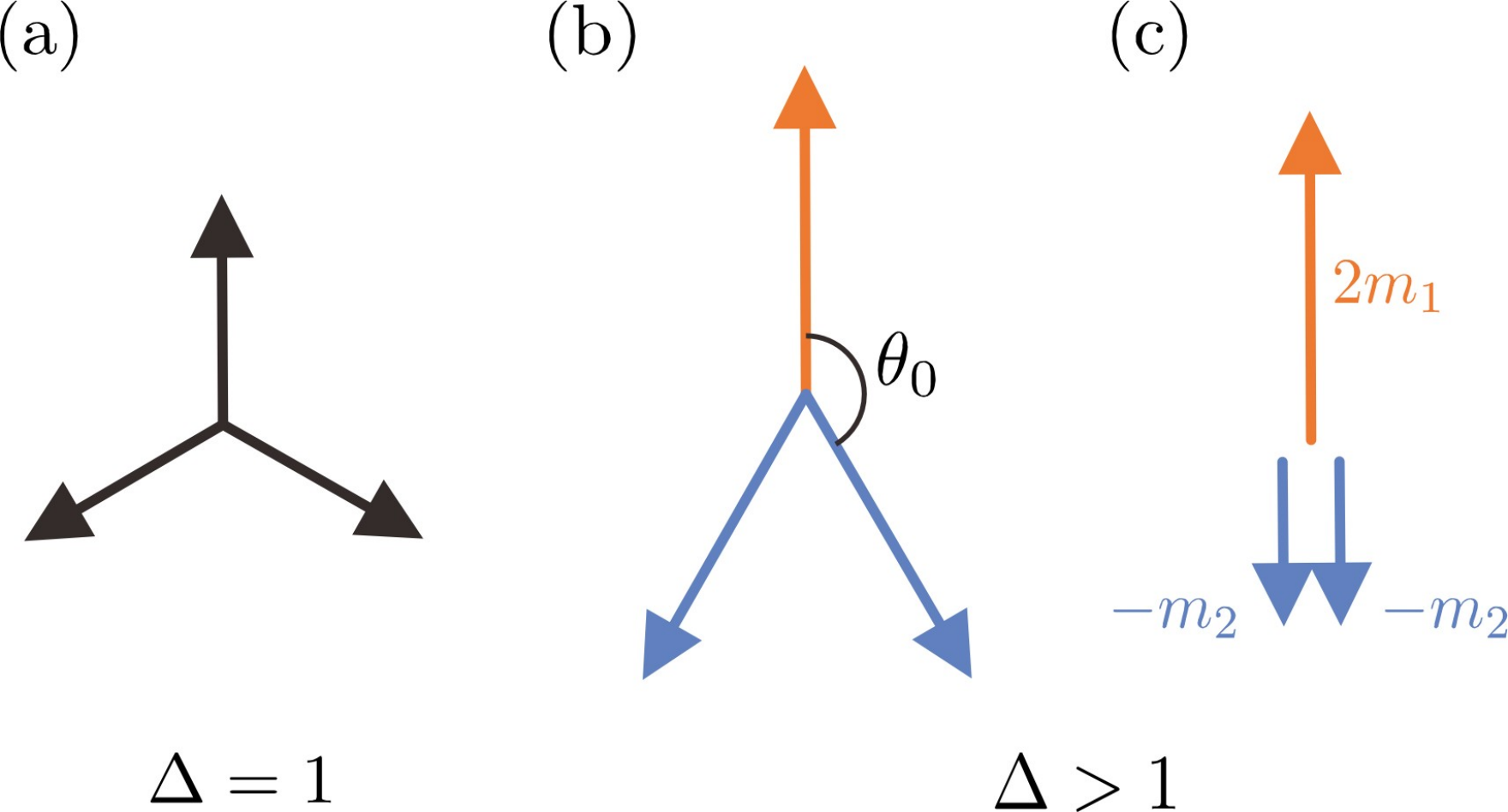}
	\caption{
		Schematic illustrations of the magnetic structures relevant to the present study.
		(a) Coplanar $120^\circ$ structure at the isotropic point $\Delta=1$.
		(b) Classical Y structure in the easy-axis region $\Delta>1$. The angle $\theta_0$ denotes the polar angle of the two tilted sublattice spins measured from the positive $z$ axis.
		(c) Three-sublattice pattern of the $z$ component of the ordered moment in the quantum case, $(2m_1,-m_2,-m_2)$.
		Panel (c) represents only the $z$ components and not the full vector spin configuration.
	}
	\label{fig:structures}
\end{figure}

Figure~\ref{fig:structures} schematically illustrates the magnetic structures relevant to the present study. At $\Delta=1$, the classical ground state is the coplanar $120^\circ$ structure [Fig.~\ref{fig:structures}(a)]. In the easy-axis region, the corresponding classical zero-field state is the Y state [Fig. 1(b)], defined above as a zero-magnetization three-sublattice coplanar state. In the present work, however, we focus on the three-sublattice pattern of the $z$ component of the ordered moment in the quantum case, schematically shown in Fig.~\ref{fig:structures}(c) as $(2m_1,-m_2,-m_2)$. This notation refers only to the $z$ components of the ordered moments and not to the full vector spin configuration.

\section{Method}
\label{sec:method}

\subsection{Spiral boundary conditions}

\begin{figure}[tbh]
	\centering
	\includegraphics[width=1.0\linewidth]{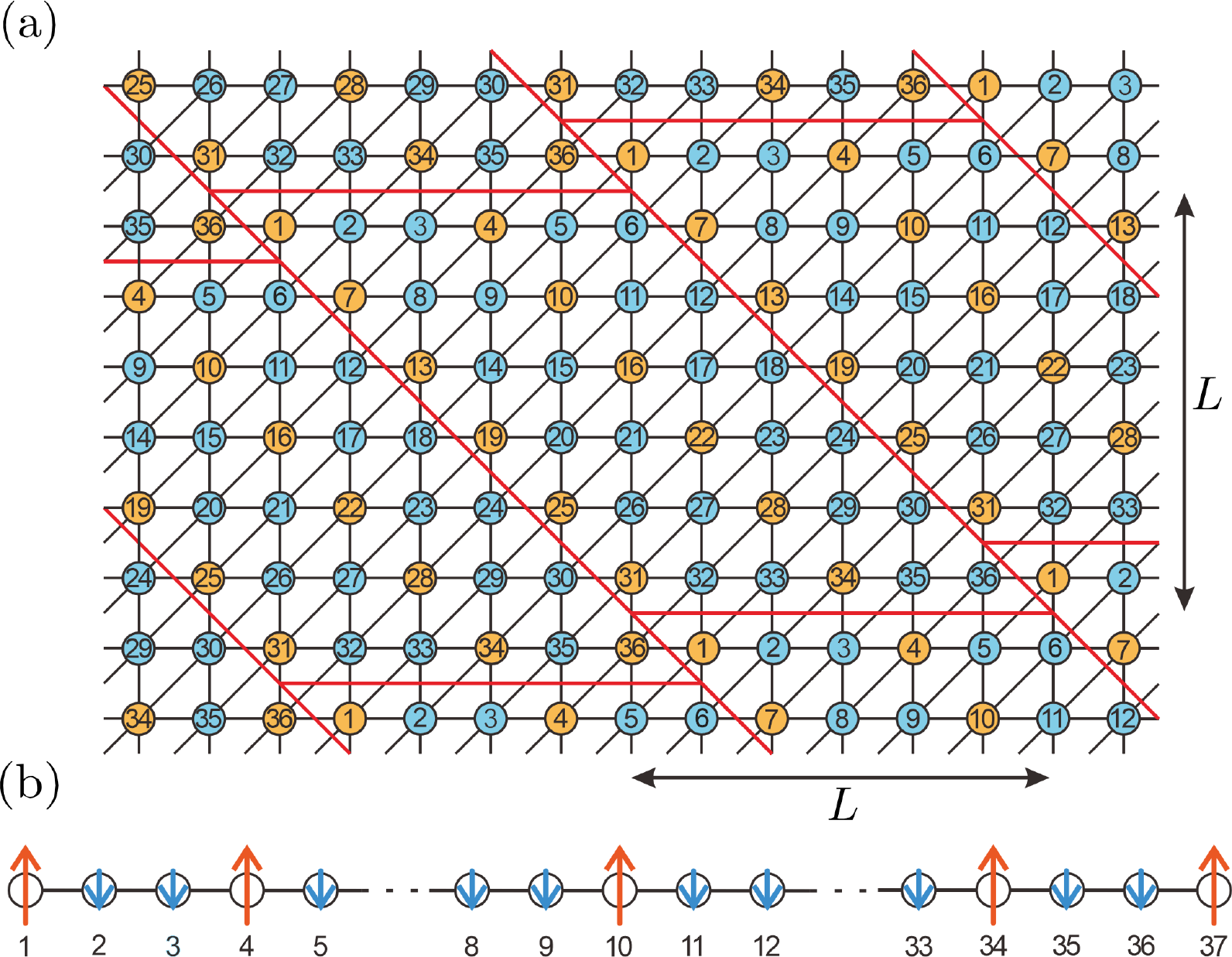}
	\caption{
		(a) Schematic illustration of the $6\times6$ triangular-lattice cluster with spiral boundary conditions. The sites corresponding to the sublattice moment $2m_1$ are shown in orange, while those corresponding to $-m_2$ are shown in blue. (b) One-dimensional mapping of the cluster in panel (a). In the actual calculation, open boundary conditions are imposed on the mapped chain. For numerical convenience, the chain is taken to contain sites up to $L^2+1$. Large pinning fields are applied to sites 1, 4, $L^2-2$, and $L^2+1$ in order to fix their spins to $S^z\approx1/2$. For visual clarity, only the nearest-neighbor bonds along the mapped chain and the pinning fields are shown in panel (b); the longer-range $(L-2)$th- and $(L-1)$th-neighbor couplings generated by the SBC mapping are omitted.
		}
	\label{fig:lattice_tri}
\end{figure}

To examine the ground-state magnetic structure, we consider finite $L\times L$ clusters of the triangular lattice. A schematic illustration of the $6\times6$ cluster is shown in Fig.~\ref{fig:lattice_tri}(a), where the sites corresponding to the positive sublattice moment $2m_1$ are shown in orange and those corresponding to the sublattices with moment $-m_2$ are shown in blue. This pattern corresponds to one of the three degenerate translational-symmetry-broken ordered states. Since the ordered state has a three-sublattice structure, the linear system size must be chosen as $L=3n$ with integer $n$.

Applying the DMRG method to two-dimensional (2D) frustrated systems is generally challenging. This difficulty originates not only from the growth of quantum entanglement with system width, but also from the fact that a straightforward one-dimensional (1D) sweeping procedure is not naturally suited to the geometry of 2D lattices. In particular, for frustrated lattices such as the triangular lattice, the choice of boundary conditions is crucial, since inappropriate finite-size geometries may artificially favor or suppress particular magnetic structures. In many DMRG studies of 2D systems, cylinder geometries are employed. However, such geometries inevitably introduce spatial anisotropy between different lattice directions, and its influence is often difficult to assess in a controlled manner.

To address this issue, we employ the SBC method~\cite{Nakamura2021,kadosawa2023-1}, which enables an exact mapping of the original 2D lattice onto a 1D chain while preserving the translational structure of the original periodic cluster. One important advantage of this method is that it avoids the spatial anisotropy inherent in cylindrical geometries. This construction also avoids artificial short bond loops that often appear in conventional cylinder or torus geometries and provides a suitable 1D representation for DMRG calculations. The corresponding 1D mapping is illustrated in Fig.~\ref{fig:lattice_tri}(b). Through this mapping, the original $L\times L$ triangular lattice is transformed into a 1D chain with nearest-, $(L-2)$th-, and $(L-1)$th-neighbor couplings, which preserves the translational structure of the original periodic cluster. The DMRG calculations are then performed on this mapped chain with open boundary conditions, as shown schematically in Fig.~\ref{fig:lattice_tri}(b). We emphasize that the $(L-2)$th- and $(L-1)$th-neighbor couplings are long-ranged only in the 1D representation; they are not additional physical interactions introduced into the model, but correspond exactly to the nearest-neighbor bonds of the original triangular lattice. Their presence increases the computational cost and makes the convergence with respect to the bond dimension more demanding, but it does not constitute an additional approximation in the Hamiltonian. In this representation, the three-sublattice pattern of the original 2D system is mapped onto a period-three order along the 1D chain.

\subsection{Density-matrix renormalization group}
\label{subsec:dmrg}

The ground state of the 1D chain obtained via SBC is investigated by means of the DMRG method~\cite{White1992}. In the DMRG calculations, we impose open boundary conditions on the projected 1D chain, which significantly improve the numerical accuracy compared with periodic chains. For numerical convenience, the site labeling is extended up to $L^2+1$, so that all sites can be indexed by a single coordinate along the 1D chain. This convention also allows the period-three ordered pattern to be arranged symmetrically with respect to the center of the 1D chain, as illustrated in Fig.~\ref{fig:lattice_tri}(b), which improves the numerical accuracy and facilitates the finite-size scaling analysis.

In the present calculations, we study the original 2D $6\times6$, $9\times9$, and $12\times12$ clusters. The lengths of the corresponding mapped 1D chains are 37, 82, and 145, respectively. We retain up to $\chi=8000$ density-matrix eigenstates and, when necessary, extrapolate physical quantities to the limit $\chi\to\infty$. The corresponding truncation error is at most of order $10^{-5}$.

To stabilize the symmetry-broken three-sublattice state in finite-size systems, we introduce local pinning fields near both ends of the open chain. The detailed choice of the pinning sites depends on the system size. As an example, for the $L \times L = 6\times6$ system, large local magnetic fields are applied to sites 1, 4, $L^2-2$, and $L^2+1$, thereby fixing their spins along the positive $z$ direction. These sites belong to the sublattice with moment $2m_1$, and, after imposing open boundary conditions, some of their bonds to other sites are removed. By contrast, sites 7 and $L^2-5$, which belong to the same sublattice but are located further inside the 1D chain, retain all the bonds present in the periodic chain and therefore do not require additional pinning fields. For larger systems such as $9\times9$ and $12\times12$, the number of pinning sites is increased accordingly.

This pinning procedure explicitly breaks the symmetry among the nearly degenerate three-sublattice configurations and selects a particular ordered state. As a result, the local magnetization profile $\langle S_i^z\rangle$ develops a clear period-three oscillation, from which the sublattice magnetization can be extracted. The role of the pinning fields is to stabilize a symmetry-broken three-sublattice profile in finite open chains, rather than to determine the value of the bulk moment. Since the sublattice moments are extracted from the central region and then extrapolated to the thermodynamic limit, the direct influence of the boundary pinning fields is expected to vanish in the final bulk estimates. We therefore regard the pinning-field calculation as a targeted calculation for extracting the bulk moments of a Y-like three-sublattice state. To determine whether this state is energetically favored, we compare its energy with that of the UDD solution and also perform calculations without pinning fields, as discussed in Sec.~\ref{subsec:energy}.

\subsection{Physical quantities}
\label{subsec:quantities}

\subsubsection{Three-sublattice magnetization}
\label{subsec:3submag}

In contrast to bipartite lattices, where the magnetic order can be characterized by the staggered magnetization, as in our previous analyses of the square- and honeycomb-lattice cases~\cite{kadosawa2023-2,Kadosawa2024}, the triangular-lattice XXZ model is expected to exhibit a three-sublattice magnetic structure. In the present analysis, we assume that the local magnetization in the easy-axis regime is described by a pattern of the form
\begin{equation}
	(2m_1,-m_2,-m_2),
	\label{eq:3sub_pattern}
\end{equation}
where the factor of 2 is introduced so that the total magnetization vanishes when $m_1=m_2$. Our aim is to determine $m_1$ and $m_2$ independently from the local magnetization profile and then examine whether they converge in the thermodynamic limit,
\begin{equation}
	(2m_1,-m_2,-m_2)\rightarrow(2m,-m,-m).
	\label{eq:3sub_limit}
\end{equation}

To this end, we calculate the local spin moment
\begin{equation}
	m_i^z=\langle S_i^z\rangle.
	\label{eq:miz}
\end{equation}
Because the pinning fields induce a clear period-three modulation of $m_i^z$ along the projected 1D chain, the sublattice moments can be extracted from the central region, where finite-size and boundary-pinning effects are expected to be smallest. The positive component is defined from the central site $i_c$ when the mapped chain has an odd length, or from a nearby site belonging to the same sublattice when no unique central site exists,
\begin{equation}
	2m_1(L,\Delta)=\langle S_{i_c}^z\rangle,
	\label{eq:m1L}
\end{equation}
while the negative component is obtained from the average over the six nearest-neighbor sites of $i_c$,
\begin{equation}
	m_2(L,\Delta)= -\frac{1}{6}\sum_{j\in {\rm NN}(i_c)} \langle S_j^z\rangle,
	\label{eq:m2L}
\end{equation}
where ${\rm NN}(i_c)$ denotes the six nearest-neighbor sites of $i_c$ on the original triangular lattice. By construction, these six sites belong to the sublattices with negative magnetization.

The thermodynamic-limit values are defined as
\begin{align}
	2m_1(\Delta) &= \lim_{L\to\infty} 2m_1(L,\Delta), \label{eq:m1inf} \\
	m_2(\Delta) &= \lim_{L\to\infty} m_2(L,\Delta). \label{eq:m2inf}
\end{align}
If the extrapolated values satisfy
\begin{equation}
	m_1(\Delta)=m_2(\Delta),
	\label{eq:m1eqm2}
\end{equation}
the ordered state is concluded to be of the form $(2m,-m,-m)$.

To confirm that this state carries zero net magnetization, we also evaluate the total magnetization per site,
\begin{equation}
	M^z(L,\Delta)=\frac{1}{N}\sum_i \langle S_i^z\rangle,
	\label{eq:Mz}
\end{equation}
and examine its size dependence. A vanishing thermodynamic-limit value of $M^z$ supports the interpretation that the easy-axis ground state is a zero-magnetization three-sublattice ordered state rather than a collinear state with finite net magnetization.

\subsubsection{Energy}
\label{subsec:energy}

As discussed in Sec.~\ref{subsec:dmrg}, the DMRG calculations with pinning fields are performed in sectors of fixed total magnetization $S_{\rm tot}^z$. Because the spin configuration near the system edges is directly constrained by the pinning fields, the total energy of a finite system is not always the most suitable quantity for characterizing the bulk magnetic structure around the center of the system.

To probe the energetics of the bulk region in the pinned calculations, we therefore evaluate the local energy around the central site $i_c$. Let ${\cal T}(i_c)$ denote the six triangles sharing the site $i_c$. We define the local energy per bond around the center as
\begin{equation}
	e_{\rm center}
	=
	\frac{1}{18}
	\sum_{t\in {\cal T}(i_c)}
	\sum_{\langle ij\rangle \in t}
	\left\langle
	S_i^xS_j^x + S_i^yS_j^y + \Delta S_i^zS_j^z
	\right\rangle,
	\label{eq:ecenter}
\end{equation}
where the sum is taken over the three edges of each of the six triangles surrounding $i_c$. Thus, shared bonds are counted more than once, and the normalization factor $18=6\times 3$ corresponds to the total number of bond contributions from these triangles. We then identify the physically relevant sector as the one that minimizes $e_{\rm center}$.

To confirm that the resulting bulk energy is not an artifact of the pinning fields, we also perform DMRG calculations without pinning fields and evaluate the total energy of the system. From this, we define the average energy per bond as
\begin{equation}
	e_{\rm tot}(L,\Delta)=\frac{E_{\rm tot}(L,\Delta)}{N_{\rm b}^{\rm 1D}},
	\label{eq:etot}
\end{equation}
where $N_{\rm b}^{\rm 1D}$ denotes the total number of bonds retained in the mapped 1D chain. The thermodynamic-limit energy obtained from $e_{\rm tot}(L,\Delta)$ is then compared with that extracted from the pinned calculations to test whether the bulk energy obtained from $e_{\rm center}$ is affected by the boundary pinning fields.

\section{Results}
\label{sec:results}

\subsection{Local $S^z$ profile}
\label{subsec:szprofile}

\begin{figure}[tbh]
	\centering
	\includegraphics[width=0.8\linewidth]{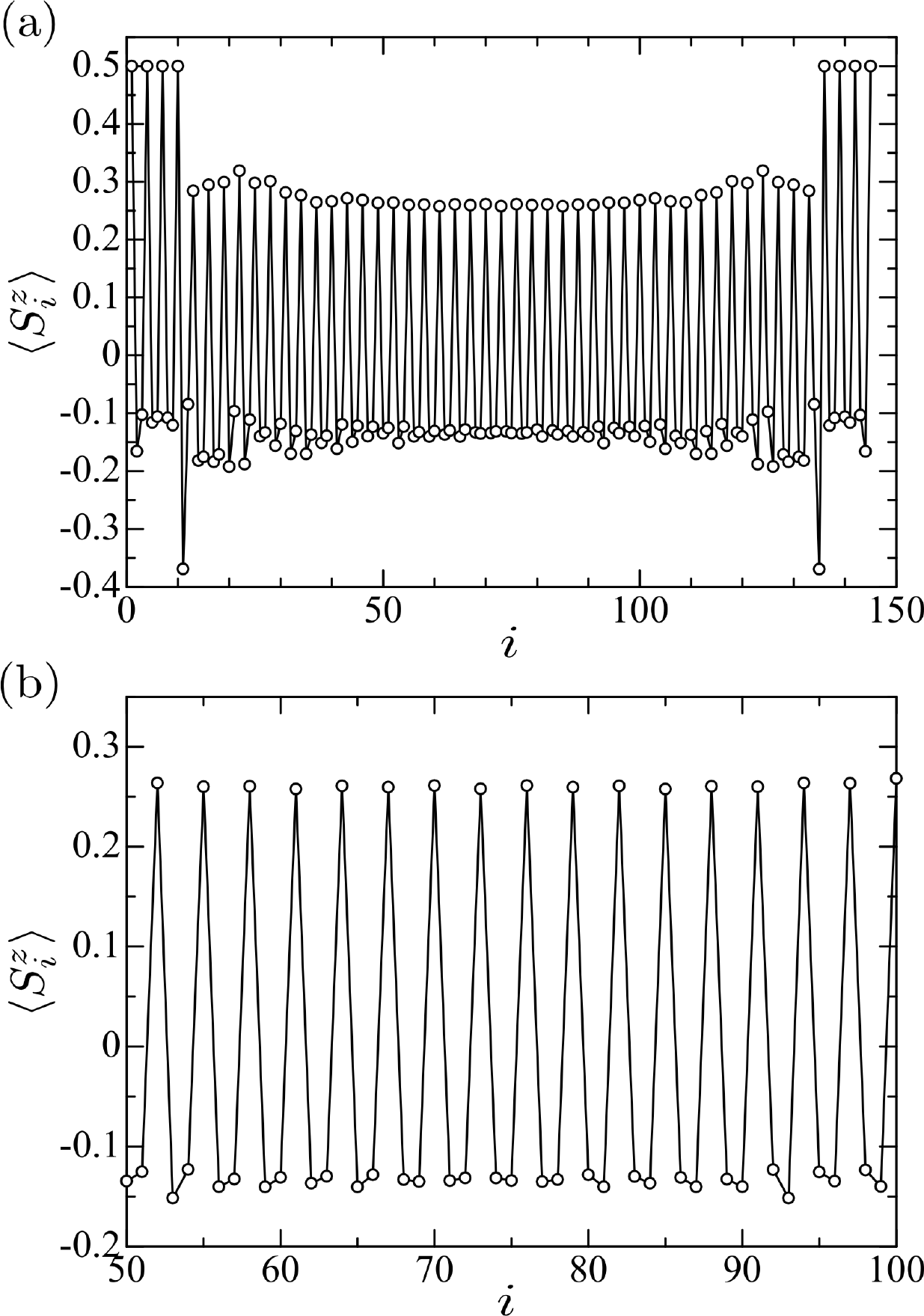}
	\caption{
		Local spin moment $\langle S_i^z \rangle$ for the projected 1D chain with 145 sites obtained from the original $12\times12$ triangular-lattice cluster at $\Delta=1.1$. Pinning fields are applied at sites $i=1,4,7,10,L^2-8,L^2-5,L^2-2$, and $L^2+1$.
		(a) Profile over the entire system, $i=1$--145.
		(b) Enlarged view of the central region, $i=50$--100. The small difference between the two negative local moments is a finite-size effect and vanishes in the thermodynamic limit.
	}
	\label{fig:Sz_profile}
\end{figure}

Before discussing the thermodynamic-limit behavior of the three-sublattice magnetization, we first examine the local spin-moment profile of a representative finite-size system. Figure~\ref{fig:Sz_profile} shows $\langle S_i^z\rangle$ for the projected 1D chain with 145 sites obtained from the original $12 \times 12$ triangular-lattice cluster at $\Delta = 1.1$.

As shown in Fig.~\ref{fig:Sz_profile}(a), the local magnetization develops a pronounced oscillatory structure over the entire system. Although the local moments near the boundaries are strongly affected by the pinning fields, the oscillation amplitude becomes nearly stationary in the central region, indicating that a well-developed bulk-like pattern is formed there. The enlarged view in Fig.~\ref{fig:Sz_profile}(b) further reveals a clear period-three modulation of $\langle S_i^z\rangle$, consistent with the three-sublattice pattern $(2m_1,-m_2,-m_2)$ introduced in Sec.~\ref{subsec:3submag}. However, in finite open chains, the two negative sublattices are not exactly equivalent because of the boundary pinning fields and the 1D representation of the cluster. Therefore, small differences between the two negative local moments can remain even near the center of the chain. We regard this asymmetry as a finite-size boundary effect rather than as evidence for a distinct bulk state with three independent sublattice moments. For this reason, the negative moment $m_2$ is defined as the average over the six nearest-neighbor sites of the central positive sublattice site, which includes both negative sublattices, and is then extrapolated to the thermodynamic limit.

These observations show that the pinning fields efficiently select a symmetry-broken three-sublattice configuration and that the central region of the finite system provides a suitable window for extracting the finite-size estimates of $m_1$ and $m_2$. We next address whether this structure survives in the thermodynamic limit by finite-size scaling.

\subsection{Estimate of the three-sublattice magnetization in the isotropic case}

As a benchmark for the present SBC+DMRG approach, we first examine the isotropic point $\Delta=1$. For the spin-$1/2$ Heisenberg antiferromagnet on the triangular lattice, the existence of three-sublattice $120^\circ$ magnetic order is well established, while quantitative estimates of the ordered moment still show noticeable method dependence~\cite{Huse1988,Bernu1992,Bernu1994,Capriotti1999,Zheng2006,White2007,Li2022,Huang2024}.

Applying the same analysis as used below for $\Delta>1$, we obtain
\begin{equation}
	2m_1 \simeq 0.217(3)
\end{equation}
for the positive sublattice moment and
\begin{equation}
	e \simeq -0.1822(7)
\end{equation}
for the ground-state energy per nearest-neighbor bond.

Our estimate of the ordered moment is somewhat larger than several recent tensor-network and DMRG estimates, but remains comparable to previous numerical results overall. For example, earlier studies reported $2m_1=0.161(5)$ from a tensor-network calculation by Li \textit{et al.}~\cite{Li2022}, and $2m_1=0.205(15)$ and $0.208(8)$ from DMRG calculations~\cite{White2007,Huang2024}. Useful summaries of previous estimates of the ordered moment are also given in Refs.~\cite{Li2022,Huang2024}. The energy is likewise comparable to previous estimates, including $e=-0.1814$ by Xiang \textit{et al.}~\cite{Xiang2001}, $e=-0.18334$ by Li \textit{et al.}~\cite{Li2022}, and $e=-0.1837$ by Iqbal \textit{et al.}~\cite{Iqbal2016}. These benchmark comparisons show that the present SBC+DMRG approach captures the isotropic three-sublattice ordered state with reasonable accuracy. Having established this reference point, we now turn to the anisotropy dependence of the three-sublattice magnetization.

\subsection{Anisotropy dependence of the three-sublattice magnetization for $S=1/2$}
\label{subsec:results_s12_delta}

\begin{figure}[t]
	\centering
	\includegraphics[width=0.8\linewidth]{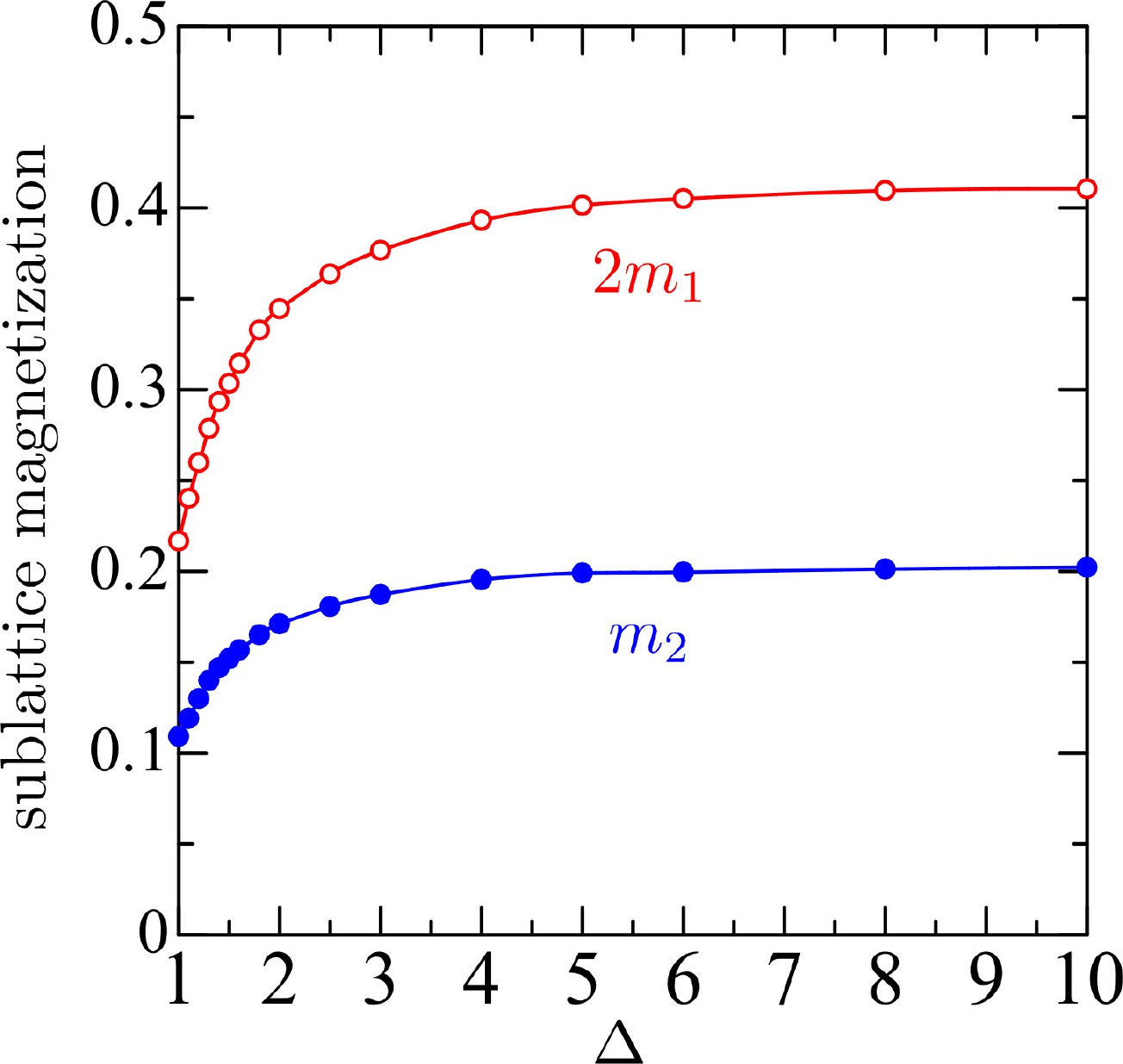}
	\caption{
		Anisotropy dependence of the positive and negative sublattice moments, $2m_1$ and $m_2$, for the spin-$1/2$ triangular-lattice XXZ model. The data are extracted from the local magnetization profile using the procedure described in Sec.~\ref{subsec:3submag}. The direct comparison between $m_1$ and $m_2$ is given in TABLE~\ref{tab:moments}.
	}
	\label{fig:m1m2_delta}
\end{figure}

We next examine the anisotropy dependence of the three-sublattice magnetization in the spin-$1/2$ case. Figure~\ref{fig:m1m2_delta} shows the $\Delta$ dependence of the positive sublattice moment $2m_1$ and the averaged negative sublattice moment $m_2$, extracted from the local magnetization profiles using the procedure described in Sec.~\ref{subsec:3submag}. Representative examples of the finite-size scaling used to obtain these quantities are presented in Appendix~\ref{app:scaling}.

\begin{table}[t]
	\caption{
		Numerical estimates of the positive and negative sublattice moments, $2m_1$ and $m_2$, for the spin-$1/2$ triangular-lattice XXZ model at several values of $\Delta$. The last column shows the relative difference between $m_1$ and $m_2$.
	}
	\begin{ruledtabular}
		\begin{tabular}{cccc}
			$\Delta$ & $2m_1$ & $m_2$ & $|m_1-m_2|/m_1$ (\%) \\
			\hline
			1.0   & 0.21671 & 0.10917 & 0.38 \\
			1.1   & 0.24020 & 0.11908 & 0.21 \\
			1.2   & 0.25990 & 0.13110 & 0.89 \\
			1.3   & 0.27860 & 0.13991 & 0.44 \\
			1.5   & 0.30343 & 0.15207 & 0.23 \\
			2.0   & 0.34449 & 0.17108 & 0.68 \\
			3.0   & 0.37658 & 0.18714 & 0.61 \\
			5.0   & 0.40144 & 0.19904 & 0.84 \\
			10.0  & 0.41049 & 0.20218 & 1.49 \\
			50.0  & 0.41668 & 0.20691 & 0.69 \\
			100.0 & 0.41753 & 0.20708 & 0.80 \\
			$\infty$ & 0.419(7) & 0.209(4) & -- \\
		\end{tabular}
	\end{ruledtabular}
	\label{tab:moments}
\end{table}

As $\Delta$ increases, both quantities evolve smoothly throughout the easy-axis regime studied here. In particular, their increase from the isotropic point $\Delta=1$ is approximately linear over the parameter range shown. A direct numerical comparison is given in TABLE~\ref{tab:moments}, which shows that $m_1$ and $m_2$ agree within about 2\% over the entire parameter range considered. This quantitative agreement indicates that the ordered state is consistently described by the form $(2m_1,-m_2,-m_2)$ with $m_1 \simeq m_2$ across a broad range of anisotropy.

These results already suggest that the ground state remains close to a three-sublattice structure of the form $(2m,-m,-m)$, rather than approaching a trivially saturated collinear Ising configuration. This interpretation is consistent with recent numerical studies, which found robust translational symmetry breaking in the easy-axis regime~\cite{Ulaga2025} and discussed the absence of a net $S^z$ magnetization in the easy-axis Y/supersolid state of the triangular-lattice $J_1$--$J_2$ model~\cite{Cesar2025}. Our results for the $J_2=0$ case, obtained using SBC, are consistent with this picture and further provide an estimate of the longitudinal three-sublattice moments in the $\Delta\to\infty$ limit. The vanishing of the net magnetization in the thermodynamic limit is examined separately in Appendix~\ref{app:net}.

\subsection{Extrapolation toward the Ising limit}
\label{subsec:results_ising}

\begin{figure}[t]
	\centering
	\includegraphics[width=0.8\linewidth]{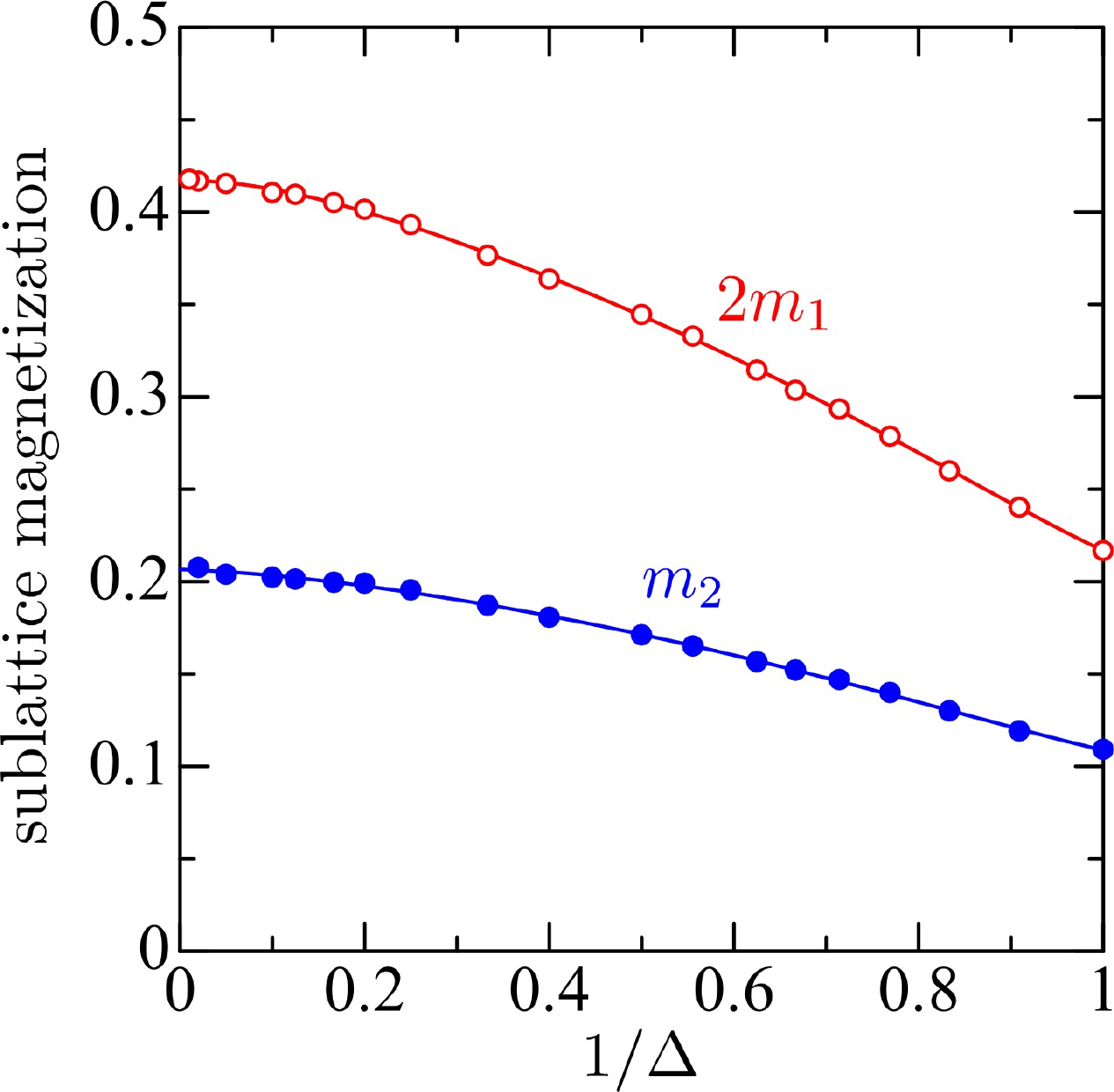}
	\caption{		
		Sublattice moments $2m_1$ and $m_2$ of the spin-$1/2$ triangular-lattice XXZ model plotted as functions of $1/\Delta$. The solid lines are empirical polynomial fits described in the text, used to extrapolate the sublattice moments toward the Ising limit, $\Delta\to\infty$. The upper edge of the vertical axis corresponds to the classical saturation value $2m_1=1/2$ of the positive sublattice moment. The extrapolated positive sublattice moment remains well below this saturation value.
}
	\label{fig:m1m2_invDelta}
\end{figure}

To examine the large-anisotropy regime in more detail, we plot $2m_1$ and $m_2$ as functions of $1/\Delta$ in Fig.~\ref{fig:m1m2_invDelta}. This representation allows us to track the behavior of the three-sublattice moments toward the Ising limit $\Delta \to \infty$.

Figure~\ref{fig:m1m2_invDelta} shows that both $2m_1$ and $m_2$ vary monotonically with $1/\Delta$. Their dependence is approximately linear over the intermediate range shown, whereas the slope gradually decreases toward the Ising limit $1/\Delta \to 0$, indicating a tendency toward saturation. Extrapolating the data to $1/\Delta \to 0$, we obtain finite values for both quantities. In this extrapolation to the Ising limit, we set $x=1/\Delta$ and fit the data for $2m_1$ and $m_2$ separately using the polynomial form $M(x)=M_\infty+\sum_{n=1}^{5} a_n x^n$, where $M_\infty$ gives the extrapolated value at $x=0$, corresponding to $\Delta\to\infty$. The polynomial is used as an empirical fitting function to describe the smooth dependence of the numerical data on $1/\Delta$, and the solid curves in Fig.~\ref{fig:m1m2_invDelta} represent these fits. In particular, the extrapolated value of the positive sublattice moment is estimated to be
\begin{equation}
	2m_1 \simeq 0.419(7),
\end{equation}
which remains significantly below the classical saturation value $1/2$. The extrapolated magnitude of the negative sublattice moment is
\begin{equation}
	m_2 \simeq 0.209(4).
\end{equation}

The extrapolated values of $m_1$ and $m_2$ are mutually consistent to within about 1\%, supporting the view that the ordered state remains close to the form $(2m,-m,-m)$ even in the large-anisotropy regime. These finite extrapolated moments therefore indicate that the easy-axis state does not approach a trivially saturated collinear Ising configuration as the Ising limit is approached from finite $\Delta$. Instead, the ordered state remains consistent with a nontrivial three-sublattice state selected from the macroscopically degenerate Ising manifold by quantum fluctuations.

We emphasize that the limiting procedure discussed here should be distinguished from the strict point $1/\Delta=0$. At exactly $1/\Delta=0$, the transverse exchange is absent and the model reduces to the classical triangular-lattice Ising antiferromagnet with a macroscopically degenerate ground-state manifold. Thus, the strict $1/\Delta=0$ point is singular, and taking the limit $1/\Delta\to0^+$ in the presence of finite transverse exchange is not equivalent to setting the transverse exchange to zero from the outset. The extrapolated moments therefore characterize the state selected by the XXZ Hamiltonian in this $1/\Delta\to0^+$ limiting procedure, rather than a unique ground state of the pure Ising Hamiltonian.

This behavior is in clear contrast to unfrustrated bipartite antiferromagnets, where the ordered moment approaches the classical Ising limit (1/2) as the anisotropy increases.

\subsection{Energy comparison between the Y state and the up-down-down state}\label{subsec:energy}

Much of the existing literature on the triangular-lattice XXZ model has focused on the role of quantum fluctuations in stabilizing the up-up-down state under an applied magnetic field~\cite{Chubukov1991,Yamamoto2014,Sellmann2015}. The zero-field problem, however, presents a distinct situation: in the strict Ising limit $\Delta=\infty$, collinear states such as the up-down-down (UDD) configuration belong to the macroscopically degenerate ground-state manifold, whereas the Y state is not itself an Ising eigenstate but emerges once finite transverse exchange lifts this degeneracy~\cite{Miyashita1986,Kleine1992,ShengHenley1992}. To our knowledge, an explicit comparison between the zero-field Y and UDD states across finite anisotropy has not been presented.

To settle this question at the classical level, we compare the zero-field energies of the two states for finite $\Delta$, obtained by directly evaluating the classical XXZ energy for the corresponding three-sublattice spin configurations. For the classical Y-state configuration, we denote by $\theta_0$ the polar angle of the two tilted sublattice spins measured from the positive $z$ axis, as illustrated in Fig.~\ref{fig:structures}(b). Minimizing the classical energy with respect to $\theta_0$ gives $\cos\theta_0=-\Delta/(\Delta+1)$, and the resulting energy per bond is
\begin{equation}
	E_Y^{\mathrm{cl}}=-\frac{\Delta^2+\Delta+1}{12(\Delta+1)},
\end{equation}
while for the collinear UDD state one obtains
\begin{equation}
	E_{\mathrm{UDD}}^{\mathrm{cl}}=-\frac{\Delta}{12}.
\end{equation}
The energy difference is
\begin{equation}
	E_Y^{\mathrm{cl}}-E_{\mathrm{UDD}}^{\mathrm{cl}}=-\frac{1}{12(\Delta+1)}<0
	\label{eq:classic}
\end{equation}
for all finite $\Delta$. Thus, at zero field, the Y state is classically favored over the UDD state as soon as the anisotropy departs from the strict Ising limit. The energy difference vanishes as $-1/(12\Delta)$ for large $\Delta$, consistent with the restoration of the Ising degeneracy at $\Delta=\infty$. We emphasize that this result is specific to the zero-field case and is distinct from the field-induced stabilization of the UDD state discussed in previous works~\cite{Chubukov1991,Yamamoto2014,Sellmann2015}.

\begin{figure}[t]
	\includegraphics[width=0.8\columnwidth]{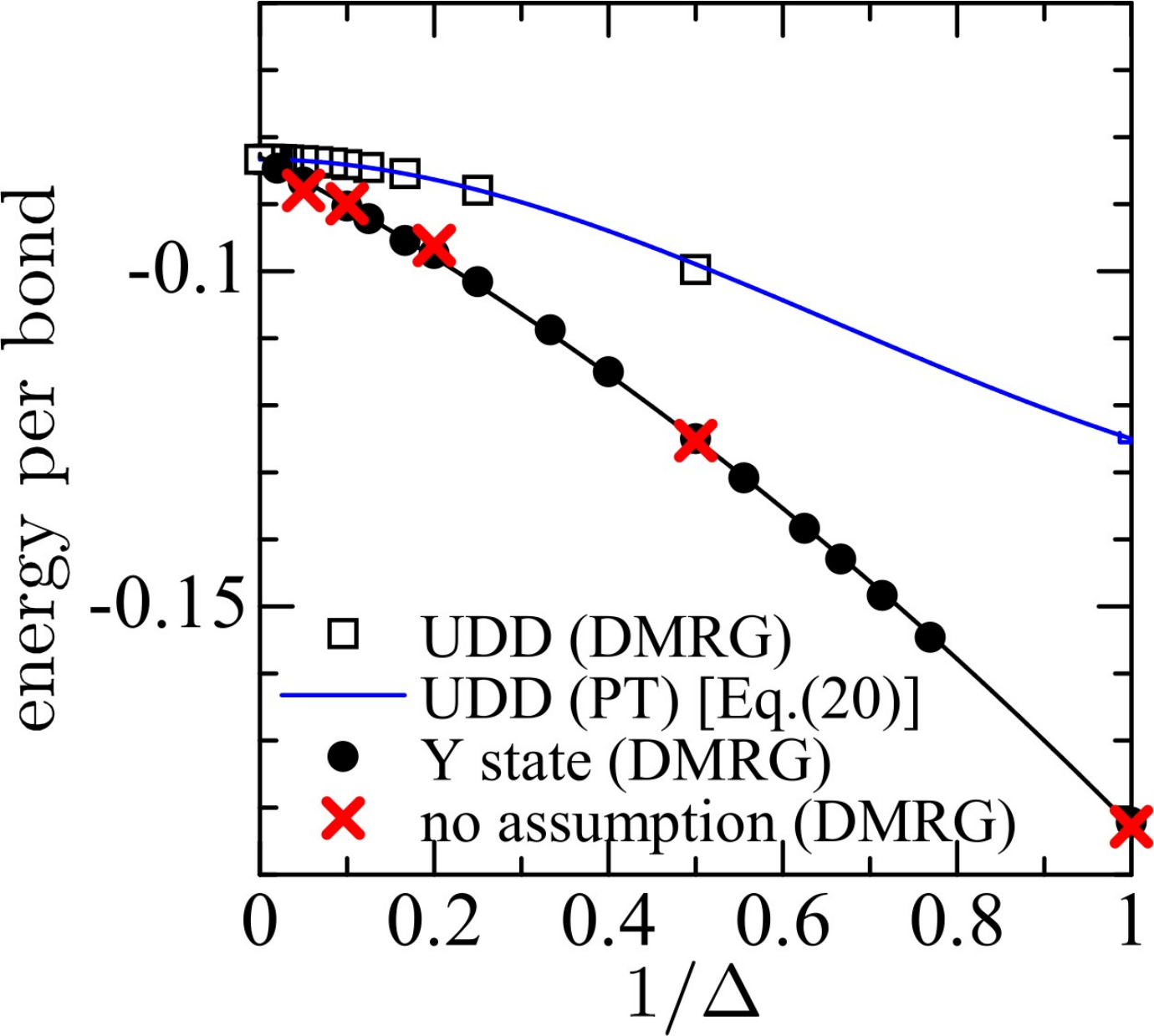}
	\caption{
		Scaled energies of the Y state and UDD state as functions of $1/\Delta$. Open squares and filled circles denote the DMRG energies of the UDD and Y states, respectively, obtained from the local energy around the center of the system ($e_{\mathrm{center}}/\Delta$), while red crosses show the thermodynamic-limit energy obtained without assuming any particular order ($e_{\mathrm{tot}}/\Delta$). The blue solid line denotes the strong-Ising expansion for the UDD state [Eq.~(\ref{eq:udd})].
		}
	\label{fig:udd_y_energy}
\end{figure}

Before turning to the quantum DMRG results, we comment on the transverse spin components. In the present finite-size DMRG calculations, the total $S^z$ quantum number is fixed and the Hamiltonian preserves the $U(1)$ spin-rotation symmetry about the $z$ axis. Therefore, the local transverse expectation values $\langle S_i^x\rangle$ and $\langle S_i^y\rangle$ vanish in finite systems and cannot be used directly as order parameters. We therefore characterize the three-sublattice structure primarily through the $z$-component of the local magnetization. The effect of the transverse exchange is nevertheless fully included in the DMRG energy. Thus, the energy comparison between the Y-state solution and the UDD solution provides a direct probe of the state selection caused by the transverse exchange at large but finite $\Delta$.

To verify this energy ordering at the quantum level, we also obtained the UDD solution directly from DMRG calculations. In this case, we used the same period-three pinning-field pattern as for the Y-state solution, and selected the UDD solution by fixing the total magnetization sector to the value compatible with the UDD configuration. Figure~\ref{fig:udd_y_energy} shows both energies as functions of $1/\Delta$. In all cases examined, the Y-state energy lies clearly below that of the UDD solution, consistent with the analytical prediction of Eq.~(\ref{eq:classic}). Moreover, the energy difference between the two solutions remains finite and well resolved throughout the parameter range studied, while decreasing toward zero in the limit $1/\Delta\to0$, in agreement with the classical result $-1/[12(\Delta+1)]$.

As an independent check on the numerical results, we also plot the perturbation-theory (PT) result for the UDD energy per bond,
\begin{equation}
	\frac{e_{\mathrm{UDD}}^{\mathrm{PT}}}{\Delta}
	=
	-\frac{1}{12}
	-\frac{1}{12\Delta^2}
	+\frac{1}{24\Delta^3}
	+O\!\left(\frac{1}{\Delta^4}\right),
	\label{eq:udd}
\end{equation}
derived in Appendix~\ref{app:pert}. As shown in Fig.~\ref{fig:udd_y_energy}, the PT result agrees well with the DMRG data in the large-$\Delta$ region, confirming the reliability of both the numerical calculations and the perturbative estimate.

More importantly, Fig.~\ref{fig:udd_y_energy} also includes the thermodynamic-limit DMRG energy obtained without assuming any particular order. The red crosses in Fig.~\ref{fig:udd_y_energy} denote the thermodynamic-limit energies obtained from calculations without pinning fields at five representative values of $\Delta$. Since the magnetic order has to develop spontaneously in the absence of pinning fields, these calculations are substantially more demanding and converge more slowly than the pinned calculations. Their agreement with the energy of the Y-state solution, together with the DMRG evidence for the $(2m,-m,-m)$ sublattice structure presented above, establishes a coherent picture: the easy-axis ground state remains Y-like rather than crossing over to a trivial collinear Ising state. The higher energy of the UDD solution further supports this interpretation by excluding the most natural collinear competitor connected to the Ising manifold. The fact that the extrapolated moment $2m\simeq0.419(7)$ remains significantly below the classical saturation value $1/2$ even as $\Delta\to\infty$ demonstrates that quantum fluctuations play an essential and nonperturbative role in this state-selection process.

\section{Conclusion}
\label{sec:conclusion}

In this work, we investigated the ground-state magnetic structure of the spin-$1/2$ triangular-lattice XXZ antiferromagnet in the easy-axis regime by means of DMRG calculations with spiral boundary conditions (SBC). As a benchmark, we first analyzed the isotropic case and confirmed that the present SBC+DMRG approach captures the three-sublattice ordered state with accuracy comparable to previous numerical estimates.

We then showed that, throughout the easy-axis region studied here, the local magnetization profile is consistently described by a zero-magnetization three-sublattice structure of the form $(2m,-m,-m)$, as supported by the close agreement between the independently estimated quantities $m_1$ and $m_2$. Furthermore, extrapolation toward the Ising limit indicates that the ordered moments remain significantly below the classical saturation value even as $\Delta\to\infty$, with $2m_1\simeq0.419(7)$ and $m_2\simeq0.209(4)$. This behavior is in clear contrast to unfrustrated bipartite antiferromagnets, where the ordered moment approaches the classical Ising limit in a straightforward manner.

We also showed analytically that the Y state is classically favored over the up-down-down (UDD) state for all finite $\Delta$ at zero field, with an energy difference of $-1/[12(\Delta+1)]$ per bond. This ordering is directly supported by DMRG calculations, which show that the Y-state energy lies below that of the UDD solution across the entire parameter range studied. More importantly, the thermodynamic-limit DMRG energy obtained without assuming any particular order agrees with the energy of the Y-state solution. Together with the persistent $(2m,-m,-m)$ sublattice structure, this establishes a coherent picture: the easy-axis ground state remains Y-like rather than crossing over to a trivial collinear Ising state.

Our results therefore demonstrate that the easy-axis ground state of the spin-$1/2$ triangular-lattice XXZ antiferromagnet is not a weakly perturbed collinear Ising state, but a nontrivial ordered state selected from the macroscopically degenerate Ising manifold by quantum fluctuations. This provides a clear example of how geometric frustration qualitatively modifies the approach to the Ising limit in quantum antiferromagnets.

\begin{acknowledgments}
We thank Ulrike Nitzsche for technical support. This work was supported by Grants-in-Aid for Scientific Research from JSPS (Projects No. JP20H01849, No. JP20K03769, and No. JP21J20604).
M.K. acknowledges support from the JSPS Research Fellowship for Young Scientists.
M.N. acknowledges the Visiting Researcher’s Program of the Institute
for Solid State Physics, The University of Tokyo, and the research fellow
position of the Institute of Industrial Science, The University of Tokyo.
S.N. acknowledges support from SFB 1143 project A05
(project-id 247310070) of the Deutsche Forschungsgemeinschaft.
\end{acknowledgments}

{\it Data availability.} The data that support the findings of this paper are openly available~\cite{data}.

\appendix

\section{Finite-size scaling of the sublattice magnetization}
\label{app:scaling}

\begin{figure}[bt]
	\centering
	\includegraphics[width=0.8\linewidth]{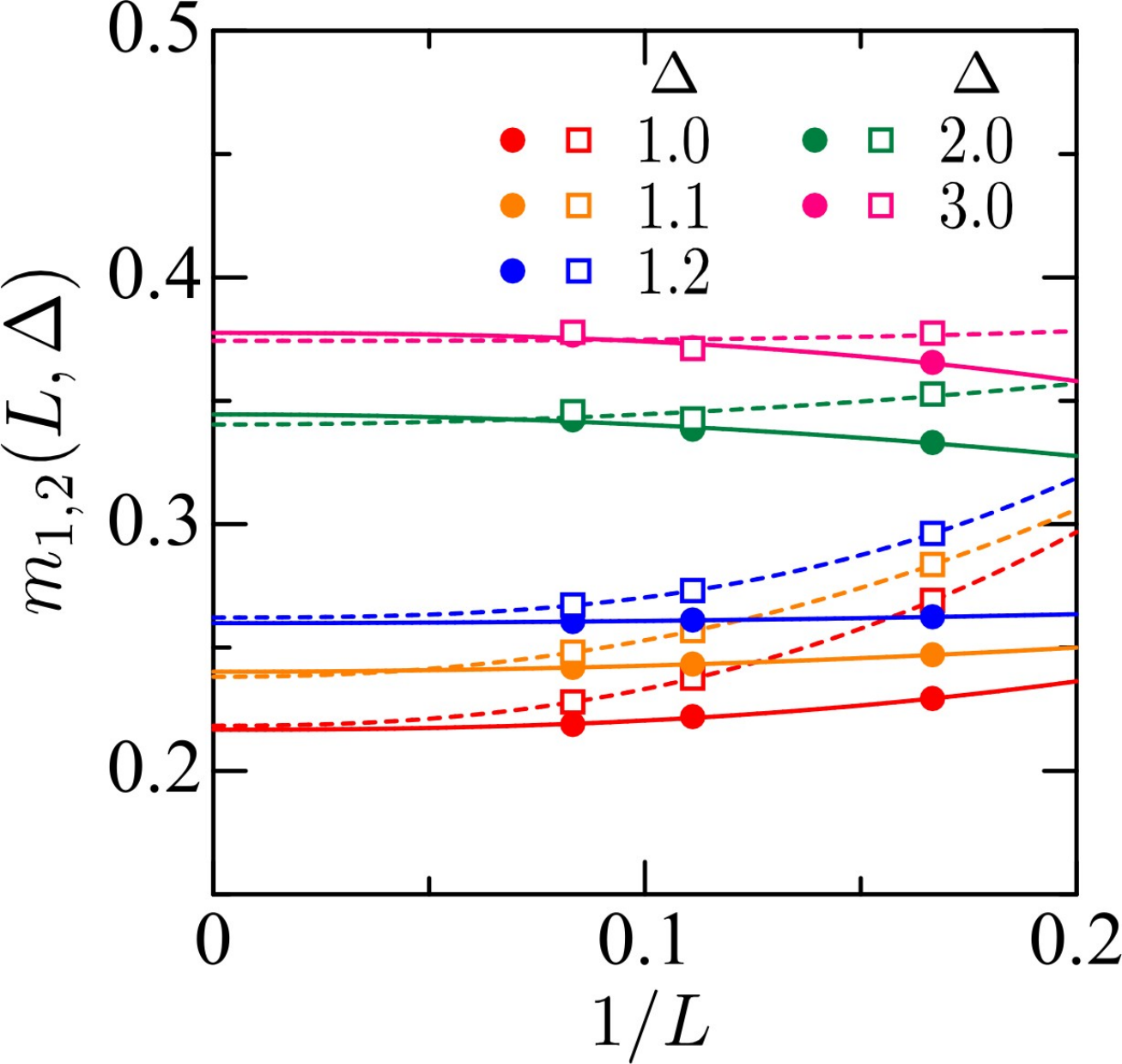}
	\caption{
		Representative finite-size scaling plots of $m_1(L,\Delta)$ and
		$m_2(L,\Delta)$ as functions of $1/L$ for several values of $\Delta$.
		The solid and dashed curves represent fits to the form
		$m(L)=a+b/L^\gamma$ for $m_1(L,\Delta)$ and $m_2(L,\Delta)$,
		respectively, used to estimate the thermodynamic-limit values.
}
	\label{fig:app_scaling_m}
\end{figure}

In this Appendix, we present representative examples of the finite-size scaling analysis for the sublattice moments introduced in Sec.~\ref{subsec:3submag}. The quantities $2m_1(L,\Delta)$ and $m_2(L,\Delta)$ are extrapolated to the thermodynamic limit using the fitting form $m_{1,2}(L)=a+\frac{b}{L^\gamma}$, where $a$, $b$, and $\gamma$ are fitting parameters. In the present analysis, the exponent $\gamma$ typically takes values between 2 and 3. The error bars quoted for the extrapolated values should be regarded as systematic uncertainties rather than statistical errors. Their dominant contribution comes from the finite-size extrapolation, and we estimate them from the variation of the extrapolated values obtained by using different reasonable fitting ranges and fitting forms.

Representative examples are shown in Fig.~\ref{fig:app_scaling_m}. In all cases displayed, both $2m_1(L,\Delta)$ and $m_2(L,\Delta)$ are well described by the fitting function, allowing stable extrapolation to the thermodynamic limit. The resulting extrapolated values are summarized in Table~\ref{tab:moments} in the main text. Their close agreement supports the conclusion that the easy-axis ground state is characterized by the three-sublattice pattern $(2m,-m,-m)$.

\section{Vanishing net magnetization in the thermodynamic limit}
\label{app:net}

\begin{figure}[t]
	\centering
	\includegraphics[width=0.8\linewidth]{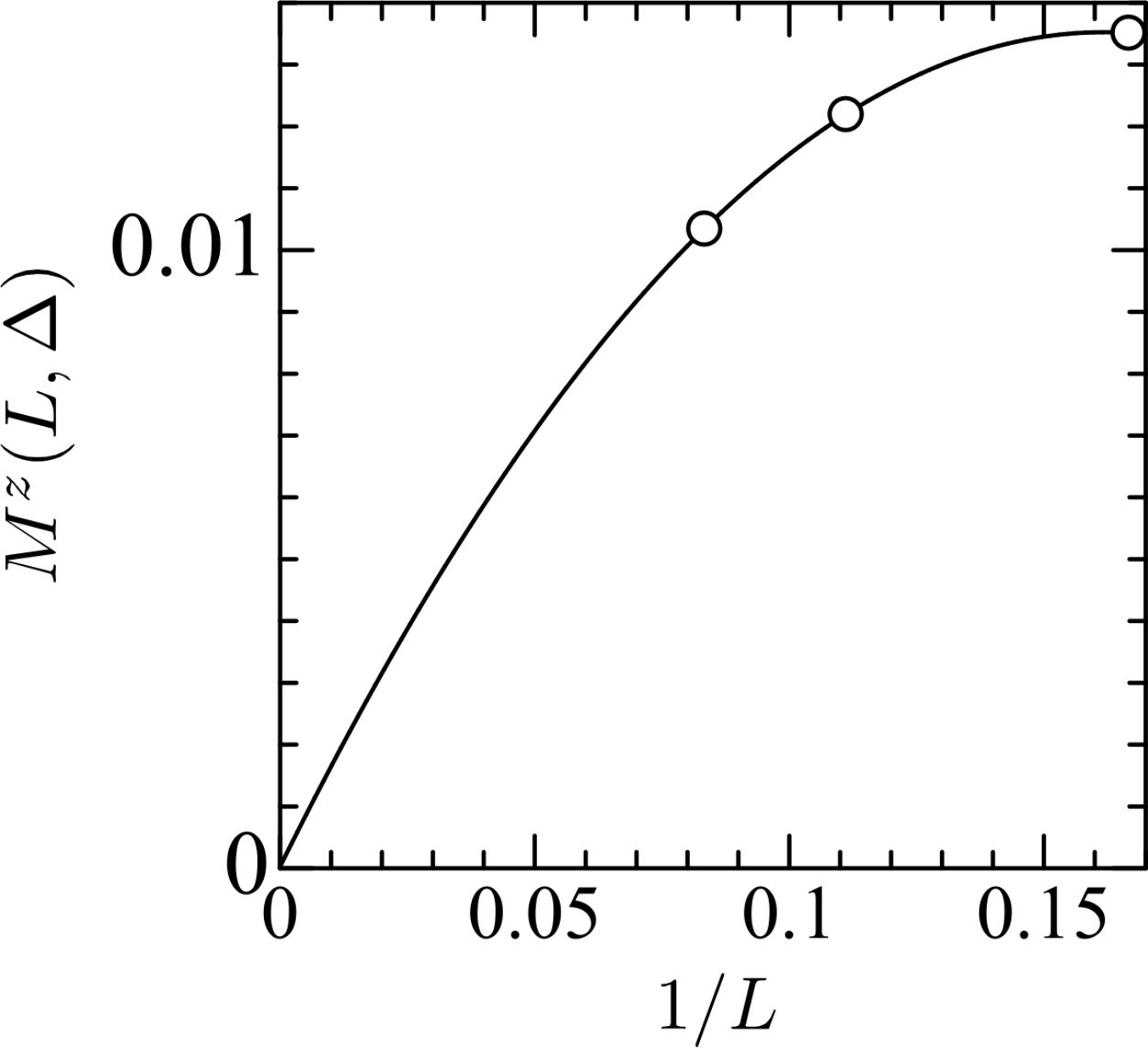}
	\caption{Finite-size scaling of the net magnetization per site
		$M^z(L,\Delta)$ as a function of $1/L$. The solid curve denotes a quadratic polynomial fit used for the extrapolation to the thermodynamic limit. The same qualitative behavior is found
		for all $\Delta$ considered.
	}
	\label{fig:Mz_scaling}
\end{figure}

In this Appendix, we examine the size dependence of the net magnetization in order to confirm that the easy-axis ground state carries zero total magnetization in the thermodynamic limit.

As discussed in Sec.~\ref{subsec:energy}, the DMRG calculations are performed in sectors of fixed total magnetization $S_{\rm tot}^z$, and the physically relevant sector is identified from the local energy around the central region of the system. In the present finite-size systems, the sector that gives the lowest energy for the Y state is found to be
\begin{equation}
	S_{\rm tot}^z=\frac{1}{2},\ 1,\ \frac{3}{2}
	\label{totalSz}
\end{equation}
for the mapped 1D chains of lengths 37, 82, and 145, respectively, corresponding to the original $6\times 6$, $9\times 9$, and $12\times 12$ clusters. Importantly, these values are selected by minimizing both the local bulk energy $e_{\rm center}$ and the average energy per bond $e_{\rm tot}$, and this identification remains unchanged throughout the entire parameter range examined, i.e., the sectors in Eq.~\eqref{totalSz} are independent of $\Delta$ within our numerical resolution. Because pinning fields are applied near the boundaries, finite systems may exhibit a small but systematic imbalance in the total magnetization even when the bulk ordered state is of the form $(2m,-m,-m)$.

This behavior can be understood heuristically as a boundary effect caused by the combination of open boundaries and boundary pinning fields. The pinning fields select the sublattice with positive $S^z$ near the two ends of the open chain, and the open boundaries make the three sublattices slightly inequivalent in finite systems. As a result, the finite clusters can acquire a small residual total magnetization even though the central bulk region realizes a nearly zero-magnetization $(2m,-m,-m)$ pattern. This imbalance is associated with the boundary rather than the bulk area, and therefore the magnetization per site decreases with increasing system size and vanishes in the thermodynamic limit.

The finite-size data of $M^z(L,\Delta)$ are plotted as functions of $1/L$ in Fig.~\ref{fig:Mz_scaling}. In the present analysis, the data are extrapolated to the thermodynamic limit by quadratic polynomial fits in $1/L$. For all values of $\Delta$ examined in the present work, the same qualitative behavior is observed: $M^z(L,\Delta)$ decreases systematically with increasing system size and is extrapolated to zero within numerical accuracy. This result provides independent support for the conclusion that the easy-axis ground state is not a UDD state with finite total magnetization, but rather a zero-magnetization three-sublattice ordered state.

\section{Perturbation theory in $1/\Delta$ around the up-down-down state}
\label{app:pert}

\begin{figure}[t]
	\centering
	\includegraphics[width=0.8\linewidth]{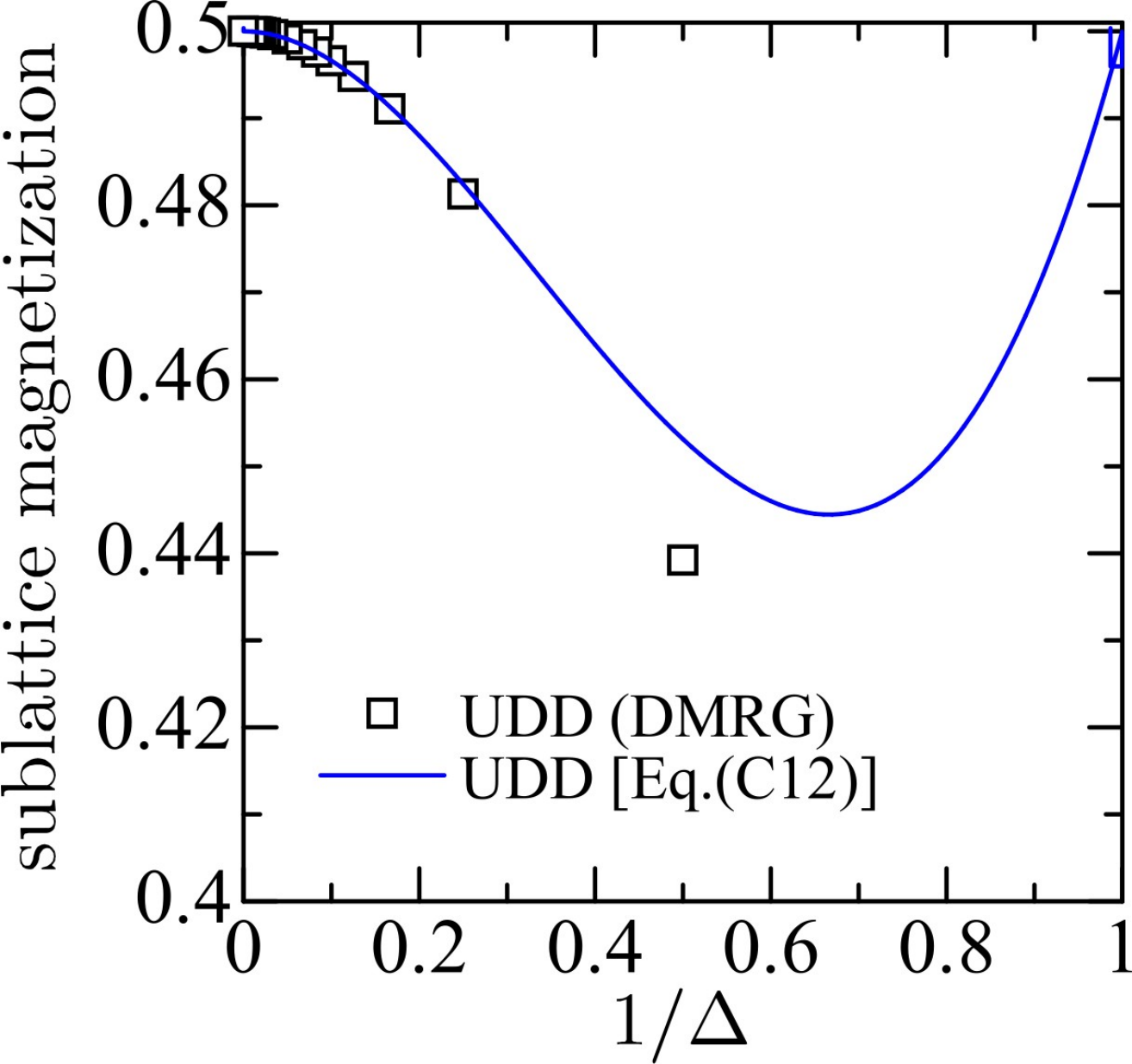}
	\caption{
		Comparison of the sublattice magnetization $m_A$ of the UDD state between DMRG and perturbation theory (PT) as functions of $1/\Delta$. The DMRG data are obtained in the magnetization sector $S^z_{\rm tot}=-N_{\rm s}/6$, while the solid curve denotes the perturbative result, Eq.~(\ref{eq:app_mA}).
}
	\label{fig:udd_mag}
\end{figure}

To benchmark the DMRG results for the up-down-down (UDD)
state, we derive strong-Ising expansions for the energy and
the sublattice magnetization in powers of $1/\Delta$. Since
the Ising limit is macroscopically degenerate, we adopt a
symmetry-broken perturbative expansion around the
three-sublattice UDD product state
\begin{equation}
	|0\rangle
	=
	\prod_{i\in A}|\!\uparrow_i\rangle
	\prod_{j\in B}|\!\downarrow_j\rangle
	\prod_{k\in C}|\!\downarrow_k\rangle .
\end{equation}
Here, $A$ denotes the up-spin sublattice, while $B$ and $C$ denote the two down-spin sublattices. The nearest-neighbor bonds of the triangular lattice are therefore classified into $AB$, $AC$, and $BC$ bonds in equal numbers. In the UDD product state, the $AB$ and $AC$ bonds are antiparallel, whereas the $BC$ bonds are parallel.

Using the Hamiltonian introduced in the main text, we
decompose it as
\begin{equation}
	H = H_0 + H',
\end{equation}
with
\begin{align}
	H_0 &= \Delta \sum_{\langle ij\rangle} S_i^z S_j^z, \\
	H'  &= \frac{1}{2}\sum_{\langle ij\rangle}
	\left(S_i^+S_j^-+S_i^-S_j^+\right).
\end{align}

The unperturbed energy per nearest-neighbor bond is obtained by this simple bond counting:
\begin{equation}
e_b^{(0)}
=\Delta\frac{(-1/4)+(-1/4)+(+1/4)}{3}
=-\frac{\Delta}{12}.
\end{equation}
The perturbation $H'$ acts only on antiparallel bonds, namely
the $AB$ and $AC$ bonds. On such a bond, $H'$ exchanges the two antiparallel spins. If $|ij\rangle$ denotes the resulting one-bond-flip state obtained from $|0\rangle$ by this exchange on the bond $\langle ij\rangle$, the nonzero matrix element is
\begin{equation}
	\langle ij|H'|0\rangle = \frac{1}{2}.
\end{equation}
The same state $|ij\rangle$ also determines the energy denominator in perturbation theory. Its unperturbed excitation energy is obtained from $H_0$ by counting the change in the Ising bond energy caused by exchanging the two spins. For example, for an $AB$ bond flip, the flipped $AB$ bond itself remains antiparallel, while the surrounding bonds connected to the two flipped sites change their parallel or antiparallel character. This bond counting gives
the excitation energy
\begin{equation}
	E_{ij}^{(0)}-E_0^{(0)} = 2\Delta.
\end{equation}
The first-order corrected wave function is therefore
\begin{equation}
	|\Psi\rangle
	=
	|0\rangle
	-\frac{1}{4\Delta}\sum_{\langle ij\rangle\in AB,AC}|ij\rangle
	+O(\Delta^{-2}),
	\label{eq:app_psi1}
\end{equation}
where the sum runs over the antiparallel $AB$ and $AC$ bonds.

Since $\langle 0|H'|0\rangle = 0$, the first-order correction
to the energy vanishes. The leading contribution therefore
appears at second order. Because the number of active bonds is
$2N_{\rm s}$ out of the total $3N_{\rm s}$ nearest-neighbor
bonds, the second-order correction to the energy per bond is
\begin{equation}
	\delta e_{\rm b}^{(2)}
	=
	-\frac{1}{12\Delta}.
\end{equation}
At third order, nonvanishing contributions arise from cyclic
exchange processes around an elementary triangle, yielding
\begin{equation}
	\delta e_{\rm b}^{(3)}
	=
	\frac{1}{24\Delta^2}.
\end{equation}
Combining these results with the unperturbed value
$e_{\rm b}^{(0)}=-\Delta/12$, we obtain
\begin{equation}
	\frac{e_{\rm b}}{\Delta}
	=
	-\frac{1}{12}
	-\frac{1}{12\Delta^2}
	+\frac{1}{24\Delta^3}
	+O(\Delta^{-4}).
	\label{eq:app_energy_scaled}
\end{equation}
Here $e_{\rm b}$ corresponds to $e_{\rm UDD}^{\rm PT}$ in the main text [Eq.~\eqref{eq:udd}].

The sublattice magnetizations follow from the corrected wave
function. From Eq.~\eqref{eq:app_psi1}, the $O(\Delta^{-2})$
terms are obtained from the norm of the one-bond-flip states.
The $O(\Delta^{-3})$ terms arise from the projection of the
second-order wave-function correction onto the same
one-bond-flip sector. For a given active bond, there are two
such second-order paths, corresponding to the two triangles
sharing that bond, which yield a coefficient
$1/(8\Delta^2)$ for each one-bond-flip state. Using the
normalized wave function, we obtain
\begin{align}
	m_A &\equiv \langle S_i^z\rangle_{i\in A}
	=
	\frac{1}{2}
	-\frac{3}{8\Delta^2}
	+\frac{3}{8\Delta^3}
	+O(\Delta^{-4}),
	\label{eq:app_mA}
	\\
	m_B &= m_C \equiv \langle S_i^z\rangle_{i\in B,C}
	=
	-\frac{1}{2}
	+\frac{3}{16\Delta^2}
	-\frac{3}{16\Delta^3}
	+O(\Delta^{-4}).
	\label{eq:app_mBC}
\end{align}
These expressions satisfy
\begin{equation}
	\frac{1}{3}\left(m_A+m_B+m_C\right)=-\frac{1}{6},
\end{equation}
as required by the conservation of the total $S^z$. One can verify that the perturbative corrections cancel order by order, as required by the conservation of the total $S^z$.

Figure~\ref{fig:udd_mag} compares the perturbative result for
$m_A$ with the DMRG data for the UDD state as functions of
$1/\Delta$. The agreement is very good in the strong-Ising
regime and remains quantitatively accurate up to
$1/\Delta \simeq 0.2$, beyond which visible deviations
gradually develop.

\begin{table}[t]
	\caption{
		Comparison between the DMRG results and the strong-Ising
		expansion for the UDD state. The DMRG data are obtained
		in the magnetization sector $S^z_{\rm tot}=-N_{\rm s}/6$.
		Here $e_{\rm b}$ denotes the energy per nearest-neighbor
		bond.
	}
	\label{tab:udd_compare}
	\begin{ruledtabular}
		\begin{tabular}{c c c c c}
			$\Delta$
			& $(e_{\rm b}/\Delta)^{\rm DMRG}$
			& $(e_{\rm b}/\Delta)^{\rm pert}$
			& $m_A^{\rm DMRG}$
			& $m_A^{\rm pert}$ \\
			\hline
			$50$
			& $-0.083366$
			& $-0.083366$
			& $\phantom{-}0.499853$
			& $\phantom{-}0.499853$
			\\
			$20$
			& $-0.083537$
			& $-0.083536$
			& $\phantom{-}0.499107$
			& $\phantom{-}0.499109$
			\\
			$10$
			& $-0.084127$
			& $-0.084125$
			& $\phantom{-}0.496589$
			& $\phantom{-}0.496625$
			\\
			$4$
			& $-0.087951$
			& $-0.087891$
			& $\phantom{-}0.481283$
			& $\phantom{-}0.482422$
			\\
			$2$
			& $-0.099786$
			& $-0.098958$
			& $\phantom{-}0.439196$
			& $\phantom{-}0.453125$
		\end{tabular}
	\end{ruledtabular}
\end{table}

In the DMRG calculations, we target the magnetization sector
$S^z_{\rm tot}=-N_{\rm s}/6$, which corresponds to the UDD
configuration. Together with the finite-size scaling of the
sublattice magnetization, this confirms that the UDD order can
be stabilized as a well-defined competing state in finite
systems. Although the Y state has a lower energy, as shown in
the main text, the UDD state nevertheless provides a useful
benchmark whose properties can be directly compared with the
perturbative results derived above.

Table~\ref{tab:udd_compare} compares the DMRG results with
the strong-Ising expansion for representative values of
$\Delta$. As expected, the agreement improves rapidly with
increasing $\Delta$.


\bibliography{SBC_triangular}

\end{document}